\documentclass[aps,prl,twocolumn,epsfig,amsfonts,amsmath,amstex,floatfix]{revtex4}

\usepackage{bm}
\usepackage{braket}
\usepackage{graphicx}
\usepackage{hyperref}

\usepackage{bbold}
\usepackage{color}

\usepackage[]{amsmath}
\usepackage{braket}
\usepackage{amsfonts,amsmath}
\usepackage{epsfig,amsmath}
\usepackage{graphicx}% Include figure files
\usepackage{dcolumn}% Align table columns on decimal point
\usepackage{bm}% bold math %\bibliographystyle{apsrev}
\usepackage{bbold}
\newcommand{\beq}{\begin{equation}}
\newcommand{\eeq}{\end{equation}}
\newcommand{\beqa}{\begin{eqnarray}}
\newcommand{\eeqa}{\end{eqnarray}}

\newcommand{\ra}{\rangle}

\begin{document}

\title{Parameter-Decoupling-Informed Learning for Superresolution: \\ Laboratory Validation and Galaxy Imaging}

\author{
Abdelali Sajia$^{1}$,
Bilal Benzimoun$^{1,2}$,
James Qu$^{1,3}$,
Pawan Khatiwada$^{1}$,
Guogan Zhao$^{1}$,
Gang Zhang$^{1}$,
and Xiao-Feng Qian$^{1}$}
\email{xqian6@stevens.edu}

\affiliation{
$^{1}$Department of Physics and Center for Quantum Science
and Engineering, Stevens Institute of Technology,
Hoboken, NJ 07030, USA\\
$^{2}$Department of Physics, Clark University,
Worcester, MA 01610, USA\\
$^{3}$Syosset High School,
70 S Woods Rd, Syosset, NY 11791, USA
}

\date{\today}

\begin{abstract}
We present a parameter-decoupling-informed learning model for estimating subwavelength separations of point sources, without requiring prior knowledge of source properties, control over the sources, or continued access to their light signals. The method builds on our parameter-decoupled quantum superresolution estimation theory, which circumvents the need to estimate nuisance parameters such as partial coherence, brightness imbalance, and relative phase. Our convolutional neural network further accommodates practical imperfections, including background noise, photon loss, and centroid and orientation misalignment. We validate the method using synthetic and experimentally generated images, achieving separation-estimation fidelity above $82\%$ for separations as small as approximately $0.075\sigma$, where $\sigma$ is the Gaussian intensity-PSF standard deviation. This corresponds to an illustrative separation of $\sim13.5$ nm when $\sigma=180$ nm, suggesting potential relevance to length scales addressed by advanced active-source microscopy. Applied to archival Hubble images of NGC~4486B, the model yields nuclear separations of 9.3-13.2~pc, consistent with the previously reported value of 12~pc. Comparisons using the same training and testing images demonstrate substantially higher fidelity than conventional direct data-driven learning in the tested cases. The method’s robustness to varying source properties and imaging imperfections supports retrospective astrophysical image analysis and motivates applications in live-cell microscopy and quantum metrology where source control or renewed optical measurements are impractical.
\end{abstract}

\maketitle

\section{Introduction}

Estimating the separation of two overlapping sources is a central task in microscopy, astronomy, and optical metrology. Diffraction broadens each point source into a point spread function (PSF), making closely spaced sources difficult to distinguish in an intensity image \cite{abbe1873beitrage,born1999principles,rayleigh1879xxxi,rayleigh1896xv,goodman2005introduction,kolobov2000quantum}. Although the Abbe and Rayleigh criteria characterize conventional optical resolution, they do not impose an absolute lower bound on the separation that can be estimated under a specified statistical model. Instead, attainable precision depends on the measurement, photon budget, noise, and available prior information \cite{bettens1999model,van2002high,ram2006beyond,chao2016fisher,stelzer2000uncertainty}. This distinction is particularly important when the objective is quantitative separation estimation rather than visual image enhancement.

%Advances in reconstruction, instrumentation, and fluorescent probes have further extended these capabilities \cite{galbraith2011super,prakash2022super,chen2023superresolution,zhao2022sparse,fang2023fluorescent}. 

Superresolution microscopy can access additional information through controlled illumination, fluorescence depletion, or emitter switching \cite{hell1994breaking,gustafsson2000surpassing,betzig2006imaging,rust2006sub,scheiderer2025minflux}. However, acquisition strategies requiring source manipulation are generally unavailable for passive sources, whose emission cannot be controlled or must remain undisturbed \cite{stelzer2002beyond}. Moreover, such strategies cannot be applied retrospectively when only a recorded image remains available.

Two complementary approaches are relevant to superresolution in these settings for passive sources. The first uses optical measurements motivated by quantum estimation theory to process the received field before detection. Spatial-mode demultiplexing can extract separation information that conventional direct imaging accesses inefficiently in the small-separation regime \cite{tsang2015quantum,tsang2016quantum,paur2016O,tsang2019resolving}. Considerable effort has addressed realistic complications, including partial coherence, brightness imbalance, multiparameter coupling, source motion, and measurement imperfections \cite{tsang2019resurgence,hradil2019quantum,hradil2021exploring,larson2018resurgence,wadood2021experimental,liang2021coherence,liang2023quantum,santamaria2023OE,hervas2024optimizing,schlichtholz2024superresolution,schlichtholz2025superresolving,napoli2019towards,sajia2022superresolution,linowski2023NJP,Rehacek2017PRA,li2024quantum,sajia2025enhanced,zhang2024superresolution,santamaria2024single}. These properties can couple to separation and complicate its inference when they are unknown. Jointly estimating the relevant nuisance parameters may require additional measurements, calibration, and statistical resources that are difficult to obtain in practice. A recently proposed parameter-decoupling approach addresses this difficulty by identifying measurement-channel contrasts that isolate separation from unknown source properties \cite{qian2026parameter}. For appropriately chosen channels, the log-probability contrast depends only on separation, while canceling brightness imbalance, coherence magnitude, and relative phase. This reduces the inference task to a single separation-dependent observable without requiring explicit estimation of those nuisance parameters. Its direct optical implementation nevertheless requires access to the incoming field. Consequently, the measurement cannot be performed retrospectively on an ordinary intensity image, such as an archival observation of a distant galaxy.

The second approach uses machine learning to infer subresolution structure and source parameters from recorded measurement data. Applications include recovery of galaxy morphology from artificially degraded survey images using generative adversarial networks \cite{schawinski2017generative}, cross-modality fluorescence superresolution \cite{wang2019crossmodality}, and reconstruction of cellular structures from single diffraction-limited frames \cite{chen2023singleframe}. Quantitative source characterization has also been demonstrated through neural-network estimation of binary-star separation and position angle from specklegrams \cite{baba1996specklegrams}. Machine learning has further been combined with optical spatial-mode measurements for reconstruction of incoherent objects \cite{frank2023O}, joint estimation of two-source separation and relative brightness \cite{gozzard2025superresolution}, and classification of strongly blurred objects in the photon-counting regime \cite{buonaiuto2025machine}. These latter methods require spatial-mode processing of the incoming field before detection, whereas image-based methods operate on intensity data already recorded by a camera. Together, these studies demonstrate the versatility of learning-assisted superresolution while highlighting the roles of training distributions, physical models, and the information retained by the acquisition process.

%The second approach uses machine learning to infer sub-Rayleigh resolution structure computationally from recorded data. Schawinski et al. used generative adversarial networks to recover galaxy morphology from artificially degraded survey images \cite{schawinski2017generative}. Wang et al. demonstrated cross-modality fluorescence superresolution \cite{wang2019crossmodality}, while Chen et al. reconstructed superresolved cellular structures from single diffraction-limited frames \cite{chen2023singleframe}. Beyond image reconstruction, Baba et al. used neural networks to estimate binary-star separation and position angle from specklegrams \cite{baba1996specklegrams}. Digital-SMLM subsequently demonstrated prediction of emitter counts and positions from unresolved fluorescence spots, including examples with approximately 30-nm emitter separation \cite{jia2024digitalsmlm}. These studies establish the potential of learning-based inference after acquisition. Their performance also highlights the importance of training distributions, reference data, and the physical information encoded in the recorded measurement.

Here, we develop parameter-decoupling-informed learning for quantitative separation estimation from recorded intensity images. The method incorporates the analytic contrast–separation relation of Ref. \cite{qian2026parameter} into a convolutional neural network (CNN), using a representation associated with same-parity Hermite–Gaussian modes. The physical relation guides the learned inference without requiring optical mode demultiplexing during application. Thus, the trained estimator operates on existing images without further access to the source light or explicit estimation of the nuisance parameters.

We assess the method using synthetic images, laboratory measurements, and archival astronomical observations. Synthetic tests examine the effects of unknown source properties, training-set size, pixel sampling, background perturbations, intensity loss, and centroid and orientation misalignment. Independent laboratory tests demonstrate separation-estimation fidelity above 0.82 for separations down to approximately $0.075\sigma$, 14 times lower than the conventional resolution limit $\sigma$, i.e., the Gaussian intensity-PSF standard deviation. Applied to archival Hubble Space Telescope images of the double nucleus of NGC 4486B, the model yields separations of 9.8–13.6 pc, consistent with the previously reported value of approximately 12 pc. Comparisons with direct image-to-separation learning using the same training and testing data further demonstrate the benefit of the parameter-decoupling-informed representation.

The central contribution is to use a nuisance-invariant physical relation to guide learning from measurements that have already been recorded. Although quantum estimation theory motivates this representation, quantum-limit attainability for an optical mode measurement does not by itself establish quantum-limited performance for an image-based estimator. Within the conditions tested here, our results demonstrate improved separation inference from overlapping source images and motivate applications in retrospective astronomical imaging, microscopy, and optical metrology.

\begin{figure*}[htbp]
\centering
\includegraphics[width=\linewidth]{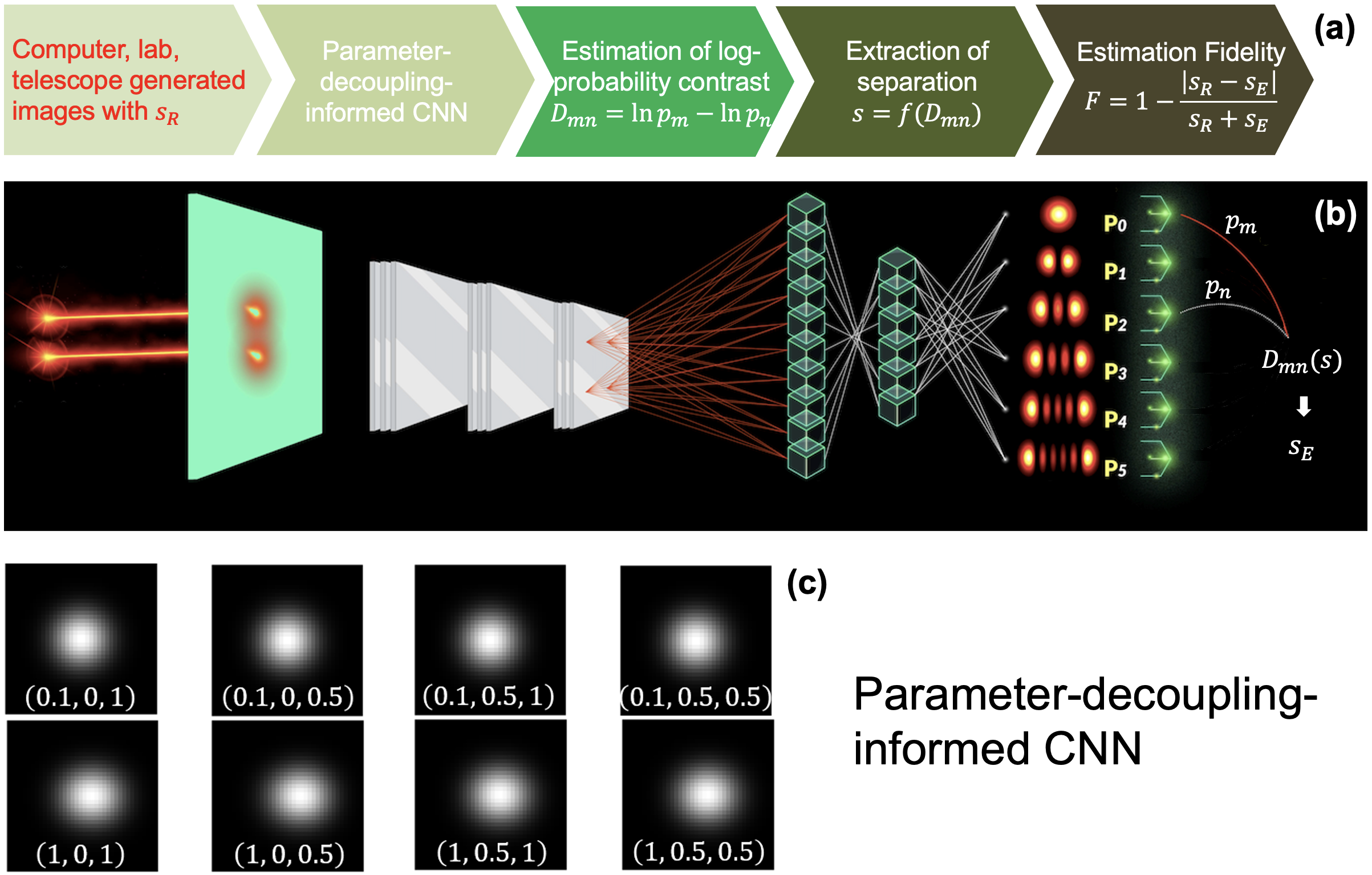}
\caption{Parameter-decoupled superresolution framework for passive sources. (a) Conceptual workflow of the parameter-decoupled protocol for estimating and evaluating two passive point-source separation. (b) Architecture of the physics-informed CNN, trained to infer the separation $s$ from input images. (c) Example training images with varying sets of parameters ($s$, $|\gamma|$, $b$) with the separation $s$ in the unit of $\sigma$. (d) Fisher information $I^{HG}$ for estimating $s$. Unlike conventional methods, $I^{HG}$ remains finite at $s=0$ (except for the idealized case of perfect destructive interference, red curve), ensuring robust superresolution across realistic experimental conditions.}
\label{roadmap}
\end{figure*}

\section{Physical model and parameter-decoupled representation}
\subsection{Two-source image model}
We describe two general partially coherent sources through the unnormalized spatial coherence operator
\begin{equation}
\begin{split}
\rho={}&p_+|h_+\rangle\langle h_+|+p_-|h_-\rangle\langle h_-|\\
&+\sqrt{p_+p_-}\bigl(\gamma|h_+\rangle\langle h_-|
+\gamma^*|h_-\rangle\langle h_+|\bigr),
\end{split}
\label{mixed state}
\end{equation}
where $h_\pm(x)=h(x\mp s/2)$ and $|h(x)|^2=\frac{1}{\sqrt{2\pi\sigma^2}}
\exp\left(-\frac{x^2}{2\sigma^2}\right)$.
Here $s\geq0$ is the separation, $\sigma$ is the standard deviation of the Gaussian intensity PSF, and $p_\pm$ specify source strengths. Without loss of generality, $p_-\geq p_+$ and $b=p_+/p_-\in[0,1]$ is the brightness ratio. The complex degree of coherence is $\gamma=g e^{i\varphi}$, with $g=|\gamma|\leq1$. The density operator is positive semidefinite, where normalization gives $\rho/\mathrm{Tr}\rho$ whenever the total detected signal is nonzero. The states overlap by $\langle h_+|h_-\rangle=\exp[-s^2/(8\sigma^2)]$. The same first-order coherence description applies to classical partially coherent fields \cite{supplemental}.

Throughout this article, small separations are reported in units of $\sigma$. This convention avoids equating the Gaussian standard deviation with an Abbe or Rayleigh distance: the Gaussian intensity full width at half maximum is $2\sqrt{2\ln2}\,\sigma$, and any conversion to a particular optical resolution criterion requires an explicit PSF convention. The laboratory results below are dimensionless separation tests, not direct demonstrations of nanometer-scale microscopy.

\subsection{Log-probability contrast}

Following Ref.~\cite{qian2026parameter}, we define the log-probability contrast
\begin{equation}
D_{\alpha\beta}=\ln p_\alpha-\ln p_\beta,
\label{eq:logcontrast}
\end{equation}
where $p_{\alpha}=\langle\phi_\alpha|\rho|\phi_\alpha\rangle$ denotes the signal probability in a spatial measurement channel $\alpha$, etc. If two channels have satisfy the general parameter-decoupling condition $p_\alpha=A(s,b,g,\varphi)f_\alpha(s)$ and $p_\beta=A(s,b,g,\varphi)f_\beta(s)$, their contrast cancels the common factor. The resulting $D_{\alpha\beta} (s)$ is independent of the nuisance parameters wherever both signals are positive.

For our machine learning consideration, here we take a specific parameter-decoupling channel solution, i.e., same parity Hermite-Gaussian modes $|\psi_\alpha\ra \equiv HG_{m0}$ and $|\psi_\beta\ra \equiv HG_{n0}$ with both $m$ and $n$ being even (or odd). Then the projected signals are given as
\begin{equation}
p_q=e^{-Q}\frac{Q^q}{q!}
\left[p_++p_-+2(-1)^q\sqrt{p_+p_-}\,g\cos\varphi\right].
\label{eq:hgprob}
\end{equation}
where $Q=s^2/(16\sigma^2)$ and $q=m,n$. As a result, the log-probability contrast simply becomes
\begin{equation}
D_{mn}
=(m-n)\ln\left(\frac{s^2}{16\sigma^2}\right)
+\ln\frac{n!}{m!}.
\label{D-contrast}
\end{equation}
Figure~\ref{roadmap}(a) labels the contrast for different pairs. Consequently,
\begin{equation}
s=4\sigma\left(\frac{m!}{n!}e^{D_{mn}}\right)^{1/[2(m-n)]}.
\label{s-D}
\end{equation}
In particular, the lowest even pair $(m=2, n=0)$ gives $s=4\sigma(2e^{D_{20}})^{1/4}$.

%In two dimensions, these channels correspond to HG$_{00}$ and HG$_{20}$ when the separation axis defines the first coordinate. Figure~\ref{roadmap}(a) labels the same ratio as $r_{02}=p_2/p_0$; Eq.~\eqref{eq:lowestpair} fixes our numerator-first convention.

These relations require a detectable contribution in the selected parity sector. At $s=0$, higher-order probabilities vanish and the logarithm must be interpreted as a limit. Fully balanced, perfectly destructive interference can eliminate the even sector altogether, making its contrast undefined. Similarly, a vanishing source strength removes an independently observable second source. Parameter cancellation does not remove these identifiability and finite-signal limitations.

According to Ref.~\cite{qian2026parameter}, separation estimation based on the log-probability contrast $D_{mn}$ asymptotically attains the nuisance-adjusted quantum Fisher information limit in the small-separation regime. This result motivates incorporating $D_{mn}$ into the learning architecture as a physically informed representation for separation estimation. Although optical quantum-limit attainability does not directly establish the performance of an image-based network, the contrast provides a principled basis for the learning model introduced below.

\section{Parameter-decoupling-informed machine learning}
\subsection{Learning representation and training data}
The CNN in Fig.~\ref{roadmap}(b) maps an intensity image to a separation estimate through a representation motivated by Eqs.~\eqref{D-contrast}--\eqref{s-D}. Instead of learning only an unconstrained correspondence between images and separation, the model uses outputs associated with the first six HG modes, $q=0,\ldots,5$, to construct same-parity contrasts. The known contrast--separation relation provides the physical structure of the inference. We use the term \emph{parameter-decoupling-informed} to distinguish this learned representation from physically measuring the HG modes.

The network uses convolutional feature extraction, average pooling, and dense layers. Six bounded outputs associated with modal contributions are used to form the contrast representation. Average pooling retains local mean intensity during downsampling; it does not select the brightest pixel. The low-order representation is motivated by the concentration of small-displacement Gaussian fields in low-order HG modes. Increasing the number of outputs is a possible extension, but is not assumed to improve estimation without a corresponding validation.

The principal synthetic dataset contains 200,000 images of $38\times38$ pixels. Separation and the source parameters $g$, $b$, and $\varphi$ are randomized in the source-uncertainty tests. Figure~\ref{roadmap}(c) shows representative profiles. The model is trained for seven epochs. Additional architecture and training procedures are referred to the Supplementary Material \cite{supplemental}. 
%The representation is tied to the Gaussian forward model and its training domain; unknown source properties do not imply an unknown PSF or arbitrary scene content.

To account for image imperfections, training includes randomly brightened and dimmed pixels, centroid displacements, and rotations. Pixel perturbations reach approximately 20\% of the 8-bit intensity scale. These augmentations are numerical robustness tests rather than complete models of detector statistics. In particular, additive pixel dimming is not equivalent to Poisson photon thinning. The first-order contrast identity cancels a common signal scale, but neither it nor intensity rescaling establishes insensitivity to arbitrary photon-number statistics at finite photon budgets.

Training was performed on a workstation with an Intel Xeon Platinum 8160 CPU at 2.10~GHz and an NVIDIA Quadro P5000 GPU with 16~GB of dedicated memory. Seven epochs required approximately 13~min in the reported implementation. This runtime characterizes the stated training task and hardware; it is not an inference-time benchmark.

\subsection{Separation-estimation metric}
We evaluate agreement between the predicted separation $s_E$ and reference separation $s_R$ using
\begin{equation}
F=1-\frac{|s_E-s_R|}{s_E+s_R}=1-\frac{|s_E/s_R-1|}{s_E/s_R+1}.
\label{fidelity}
\end{equation}
For nonnegative separations with $s_E+s_R>0$, $0\leq F\leq1$, $F=1$ indicates exact agreement, and $F=0$ suggests maximum disagreement with  $s_E$ dramatically off with the ground truth number, i.e,. $s_E \gg s_R$ or $s_E \ll s_R$. We call $F$ the separation-estimation fidelity, and it is not quantum-state fidelity or a classification success probability. All fidelity results below refer to held-out images or laboratory measurements, and the galaxy application lacks a known truth value $s_R$ and is evaluated by comparison with literature instead.

%If both separations vanish, Eq.~\eqref{fidelity} is undefined and requires a separately specified convention.

%For positive separations, $F=2\min(s_E,s_R)/(s_E+s_R)$. Accordingly, $F\geq0.8$ permits $2/3\leq s_E/s_R\leq3/2$, while $F\geq0.82$ permits approximately $0.695\leq s_E/s_R\leq1.439$. The threshold is a useful common comparison criterion, not a universal definition of resolution. 

\section{Separation estimation from synthetic images}

We first test our parameter-decoupling-informed model with synthetic images. It is designed to examine source-parameter variation, image imperfections, and training-set properties. Their ground-truth separations are known from generation, allowing comparison through Eq.~\eqref{fidelity}.

\begin{figure*}[htbp]
\centering
\includegraphics[width=0.80\linewidth]{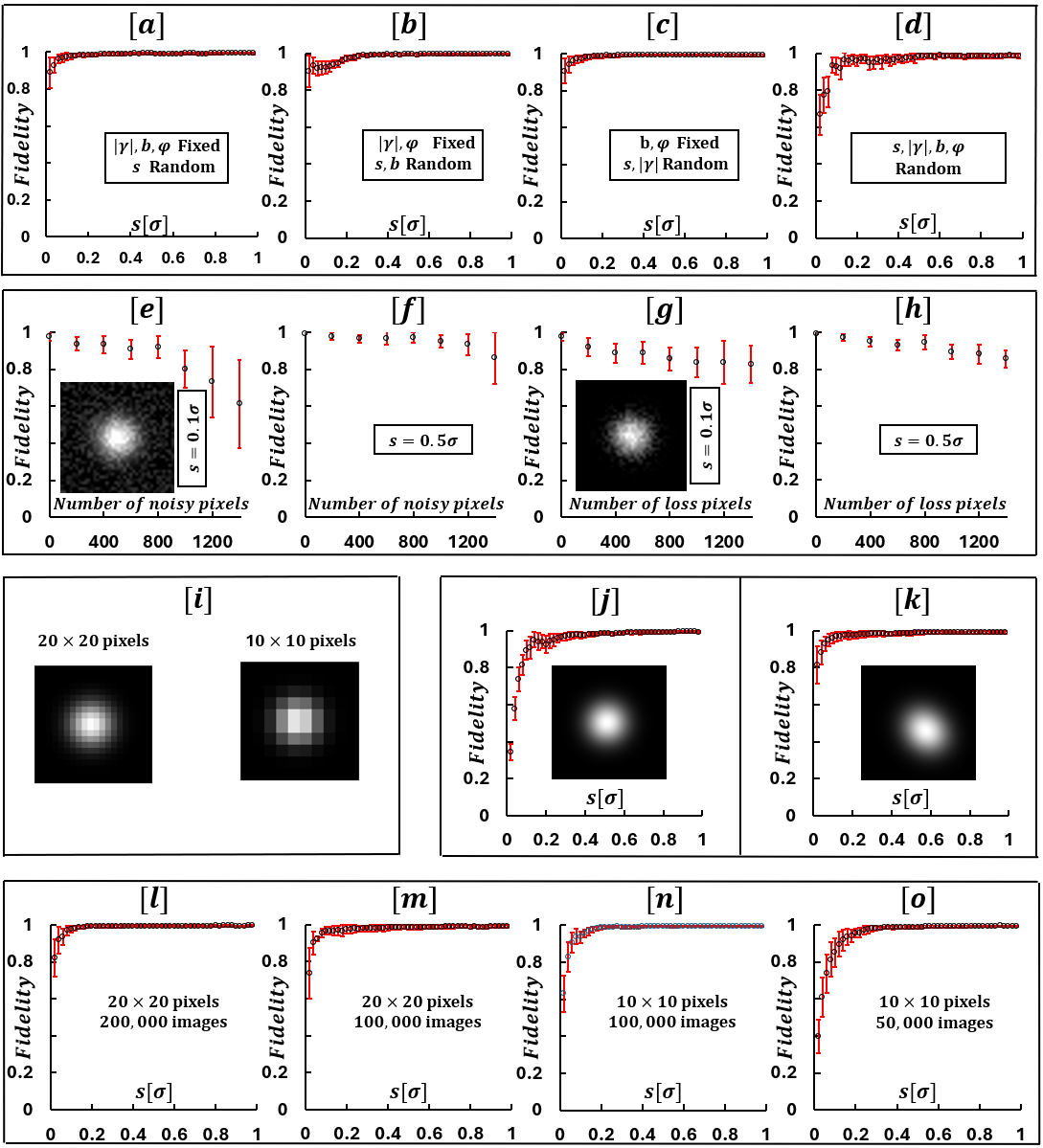}
\caption{Separation estimation fidelity of the physics-informed CNN model for computer-generated data. Effects of source-dependent parameters: (a) illustrates the case when only $s$ is unknown while other parameters are fixed $\gamma=0$, $b=1$, $\varphi=0$, (b) represents when $s$ and $b$ are unknown, while $\gamma=0$, $\varphi=0$ are fixed, (c) shows the case when both $s$ and $\gamma$ are unknown, while $b=1$ and $\varphi=0$ are fixed, and (d) depicts when all four parameters are unknown and randomized. Effects of source-independent factors: (e) and (f) illustrate the fidelity with respect to the number of polluted pixels with luminosity noise for separations $s=0.1\sigma$ and $0.5 \sigma$ respectively. A sample image with 800 noisy pixels is shown for separation $s=0.1\sigma$. Panels (g) and (h) illustrate the fidelity with respect to the number of polluted pixels with photon loss for separations $s=0.1\sigma$ and $0.5 \sigma$ respectively. A sample image with 1000 affected pixels is shown for separation $s=0.1\sigma$. Panel (j) illustrates the fidelity with respect to $s$ for data images with 5\% centroid misalignment and a sample data image, and panel (k) illustrates fidelity with respect to $s$ for data images with arbitrary orientation misalignment and a sample data image. Effects of training properties: Panel (i) illustrates two data images for separation $s=0.1\sigma$ with low resolutions of ($10\times10$) and ($20\times20$) pixels, respectively. Panels (l-n) illustrate fidelity behaviors for different number of training images and different number of pixels per image.}
\label{multiple-parameter noise}
\end{figure*}

\subsection{Unknown source parameters}
Figures~\ref{multiple-parameter noise} (a)-(d) illustrates the fidelity of our model with respect to the two-source separation for testing images with progressively increase source uncertainty.

Panel (a) illustrates fidelity of testing images by fixing nuisance parameters at $g=0$, $b=1$, and $\varphi=0$ with only the separation $s$ varying. Panel (b) additionally randomizes $b$, and panel (c) instead randomizes $g$ and $s$ with fixed parameters $b=1$, $\varphi=0$. Across these three configurations, the reported mean fidelity is approximately $F=0.9$ at $s=0.02\sigma$ (i.e., 50 times lower than conventional resolution limit $\sigma$) and approaches unity as separation increases.

Panel (d) presents the most general case by randomizing all four parameters for the testing images. In this more demanding test, the reported fidelity exceeds 0.82 for $s\gtrsim0.06\sigma$. This is separation estimation at approximately one-sixteenth of the resolution limit (Gaussian PSF standard deviation), within the simulated distribution. The increased difficulty at small $s$ is consistent with overlapping profiles becoming less distinguishable and nuisance-parameter effects becoming more consequential.

\subsection{Image imperfections}
We also test the performance of our CNN model for imperfect testing images, simulating realistic situations. To isolate image-corruption effects, we fix the nuisance parameters $g=0$, $b=1$, $\varphi=0$. Pixel intensities are represented on the scale $[0,255]$. 

Background-like corruption is generated by adding a random value between 10 and 50 to randomly selected pixels before rescaling. The number of affected pixels increases in steps of 200. These perturbations span approximately 4\%--20\% of the full 8-bit scale; their strength relative to the local signal can be larger in dim pixels. Figures~\ref{multiple-parameter noise} (e) and (f) show results for $s=0.1\sigma$ and $0.5\sigma$. At $s=0.1\sigma$, the reported fidelity remains approximately 0.9 with 800 affected pixels, or 55\% of the total pixels in a $38\times38$ image. At $s=0.5\sigma$, comparable performance extends to approximately 1200 affected pixels, or 83\%. The example inset in panel (e) illustrates the visibly corrupted profile.

For dimming tests, a random value between 10 and 50 is subtracted from randomly selected pixels, with negative intensities clipped to zero. Figures~\ref{multiple-parameter noise} (g) and (h) show the corresponding fidelities. At $s=0.1\sigma$, fidelity remains approximately 0.9 up to 600 affected pixels, or 42\% of the total number of pixels. At $s=0.5\sigma$, comparable performance extends to approximately 1000 pixels, or 69\% of the pixels. 

These results apparently demonstrate the strong reliability of the parameter-decoupling-informed CNN model in estimating two-source separations against noise and photon detection inefficiency. In particular, the relatively small error bar highlights the stability of our model.

%These results demonstrate tolerance to the specified perturbations; a photon-counting detector model would require additional tests with calibrated counting and readout noise.

Panel (j) examines a one-pixel centroid displacement, described in the original test convention as a 5\% shift. Training includes randomized centroid offsets described as 7.5\%. The estimator retains mean fidelity above 0.8 for $s\gtrsim0.08\sigma$ in this test. Panel (k) examines rotations over an angular range up to $\pi/2$, with performance close to the aligned case. Together, these tests assess tolerance to position and orientation changes without requiring their values as inference inputs. They do not independently quantify centroid-localization or orientation-estimation precision.

\subsection{Training-set size and pixel sampling}
Figures~\ref{multiple-parameter noise} (l)-(o) compare image sampling and training dataset size with $g=0$, $b=1$, $\varphi=0$, and no added image corruption. With 200,000 training images of $20\times20$ pixels, panel (l) gives mean fidelity above 0.9 for $s\gtrsim0.04\sigma$ (20 times lower than resolution limit $\sigma$). With 100,000 images, the $20\times20$ and $10\times10$ configurations in panels (m) and (n) both reach this level at approximately $0.06\sigma$ (17 times lower than $\sigma$). With 50,000 images of $10\times10$ pixels, panel (o) reaches it at approximately $0.12\sigma$ (8 times lower than $\sigma$). Example training images of the two pixel-size cases appear in panel (i).

At fixed sampling, larger training sets improve the reported small-separation performance. At fixed training-set size, increasing the pixel count does not necessarily yield a comparable gain. These observations show a practical interaction between sampling and training coverage. They do not establish the number of training images per pixel as a universal scaling law, because resampling also changes the image representation and the features available to the network.

\section{Laboratory validation}
\begin{figure*}[htbp]
\includegraphics[width=0.80\linewidth]{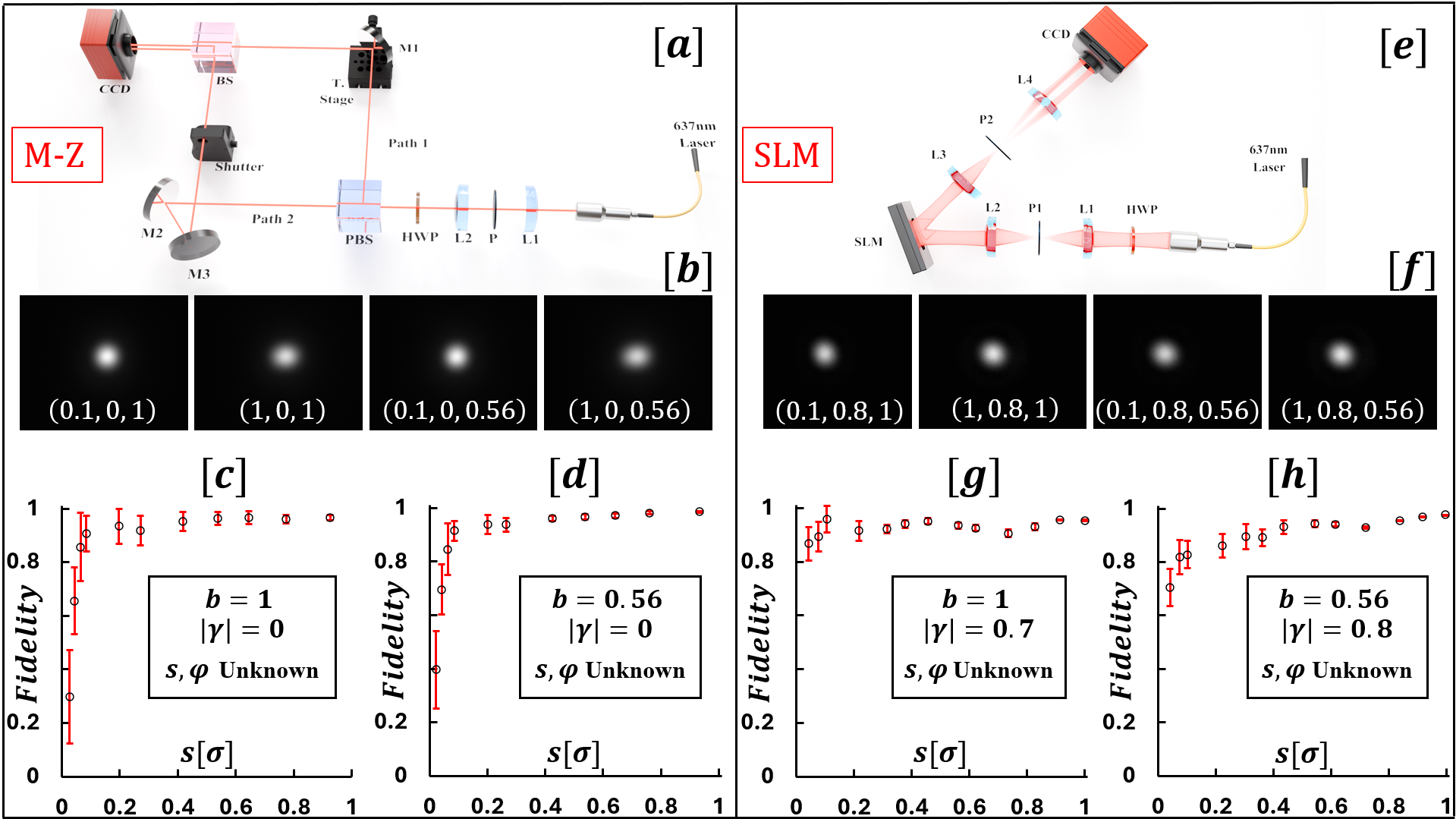}
\caption{Separation estimation fidelity of the physics-informed CNN model for experimental data. MZ Interferometer Experiments: (a) schematic illustration of the MZ interferometer setup to generate two-point source data images; (b) experimental raw data images for varying sets of parameters ($s$, $|\gamma|=0$, $b$), with fixed phase $\varphi=0$; (c) and (d) illustrate fidelity behaviors for incoherent ($|\gamma|=0$) data under balanced ($b=1$) and partially balanced ($b=0.56$) conditions, respectively. SLM Setup Experiments: (e) schematic illustration of the SLM setup to generate two-point source data images; (f) experimental raw data images for varying sets of parameters ($s$, $|\gamma|=0$, $b$) with fixed phase $\varphi=0$; (g) shows the fidelity behaviors for balanced ($b=1$) and partially coherent ($|\gamma|=0.7$) SLM testing data; (h) illustrates the fidelity behaviors for partially balanced ($b=0.56$) and partially coherent ($|\gamma|=0.8$) SLM testing data.}
\label{experiment}
\end{figure*}

We assess transfer from simulated testing to laboratory images using a Mach-Zehnder (MZ) interferometer and a spatial light modulator (SLM), shown in Figs.~\ref{experiment} (a) and (e) respectively. These setups generate controlled two-source profiles with independently calibrated separations and selected coherence and brightness ratios. Although the illumination is prepared in the laboratory, the estimator receives only recorded camera images. Source-generation hardware is not part of its computational inference stage. Setup details are referred to the Supplementary Material \cite{supplemental}.

Reference separations are obtained by localizing the source contributions individually. The MZ configuration uses a shutter and translation stage, while the SLM uses computer-generated holograms. These procedures establish the laboratory ground truth separations used in Eq.~\eqref{fidelity}. They are distinct from the simultaneous two-source image supplied to the CNN. Before inference, the images are cropped around the signal and resized to $38\times38$ pixels. The raw testing images in Figs.~\ref{experiment} (b) and (f) illustrate differences in acquisition geometry and sampling between the two setups.

Figures~\ref{experiment} (c) and (d) show incoherent ($\gamma=0$) MZ-generated images with $b=1$ and $b=0.56$ respectively. Both configurations give reported fidelities above 0.85 for $s\gtrsim0.064\sigma$ (approximately 16 times lower than the resolution limit). The SLM tests extend the assessment to partial coherence. For $g=0.7$ and $b=1$, panel (g) gives fidelity above 0.87 for $s\gtrsim0.04\sigma$ (approximately 25 times lower than the limit $\sigma$). For $g=0.8$ and $b=0.56$, panel (h) gives fidelity above 0.82 for $s\gtrsim0.075\sigma$ (approximately 1/13 of the resolution limit).

These results demonstrate high-fidelity estimation from experimentally acquired overlapping profiles under the tested source conditions. Differences between the MZ and SLM curves may reflect calibration uncertainty, aberrations, background, and sampling. They may also contain different physical noise effects. Thus these tests do not constitute independent validation over every possible combination of photon statistics, coherence, and detector imperfections. Their main role is to establish that the learned representation remains useful beyond purely synthetic input images.

\section{Application to archival imaging of NGC~4486B}
Galactic nuclei provide an application in which source fields may not be immediately ready for measurement and only previously recorded data may be available. Their morphology constrains the central stellar distribution and contributes to studies of galaxy evolution \cite{kormendy2004secular}.
Lauer \textit{et al.} identified the double nucleus of NGC~4486B in Hubble Space Telescope (HST) Wide Field and Planetary Camera~2 observations \cite{lauer1996hubble}. Their analysis used 40 iterations of Richardson--Lucy deconvolution \cite{richardson1972bayesian,lucy1974iterative} and reported a peak separation of approximately 0.15~arcsec, corresponding to 12~pc at the adopted distance \cite{lauer1996hubble}. However, because the two nuclei lie close to each other and near the telescope’s resolution limit (0.046 arcsec/pixel for the planetary camera (PC1) of HST WFPC2), their precise separation could not be reliably determined. This motivates us to use our parameter-decoupling-informed CNN method to extract more accurate structural information from this double-nucleus system.

\begin{figure*}[htbp]
\centering
\includegraphics[width=0.80\linewidth,trim=0 35 0 0,clip]{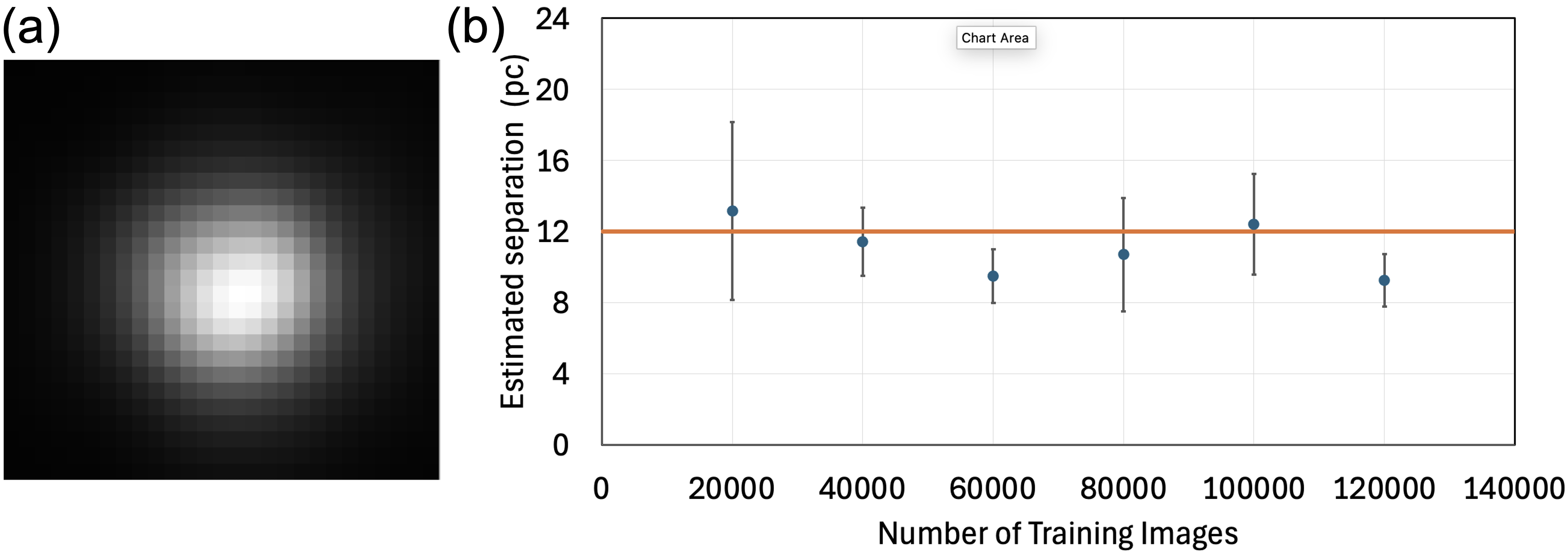}
\caption{(a) Original HST image of the NGC 4486B double-nucleus system. (b) Processed $33\times33$-pixel version of the object. (c) Double-nuclei separation data achieved with physics-informed CNN.} \label{fig:exper}
\end{figure*}

We apply the learning approach to an HST image of NGC~4486B obtained from the Hubble Legacy Archive. Following the galaxy-analysis procedure, the FITS image is converted to JPEG, a $66\times66$-pixel region around the nucleus is cropped, and the region is downsampled to $33\times33$ pixels. Intensities are then rescaled to $[0,255]$. Figures~\ref{fig:exper} (a) and (b) show the original image and processed nuclear region. The analysis adopts the galaxy distance of approximately 17~Mpc from Earth. Recent HST/JWST work provides additional context for the central structure and HST image sampling \cite{tahmasebzadeh2025jwst}. The angular scale must be propagated through the downsampling before conversion to parsecs as resampling changes the physical scale per output pixel.

For this application, eight training-set configurations contain between 10,000 and 110,000 synthetic images of matching pixel dimensions. Thus, this demonstration uses an application-specific input format and training study, rather than claiming direct transfer of the unchanged $38\times38$ laboratory model.

For configurations containing more than 10,000 training images, the reported mean separation estimates range from 9.3 to 13.2~pc, see Fig.~\ref{fig:exper}(c). This range encompasses the earlier approximately 12~pc value. Larger training sets generally reduce the displayed dispersion, indicating improved stability within this training study. 

The comparison supports consistency between the learned estimate and a previously reported analysis of the same object. It does not provide independent ground truth or demonstrate greater accuracy than Richardson--Lucy deconvolution. Unlike the laboratory profiles, galactic nuclei are extended and embedded in structured emission. A Gaussian two-source model is therefore an effective approximation whose PSF width, background treatment, pixel scale, and adopted galaxy distance affect the physical separation. More detailed source and telescope models, including diffraction-based PSFs, would be needed to establish the corresponding systematic uncertainties. The application nevertheless demonstrates that parameter-decoupling-informed learning can be used retrospectively on astronomical image data without renewed access to the source light.

\section{Comparison with direct image-to-separation learning}
\begin{figure*}[htbp]
\centering
\includegraphics[width=0.80\linewidth]{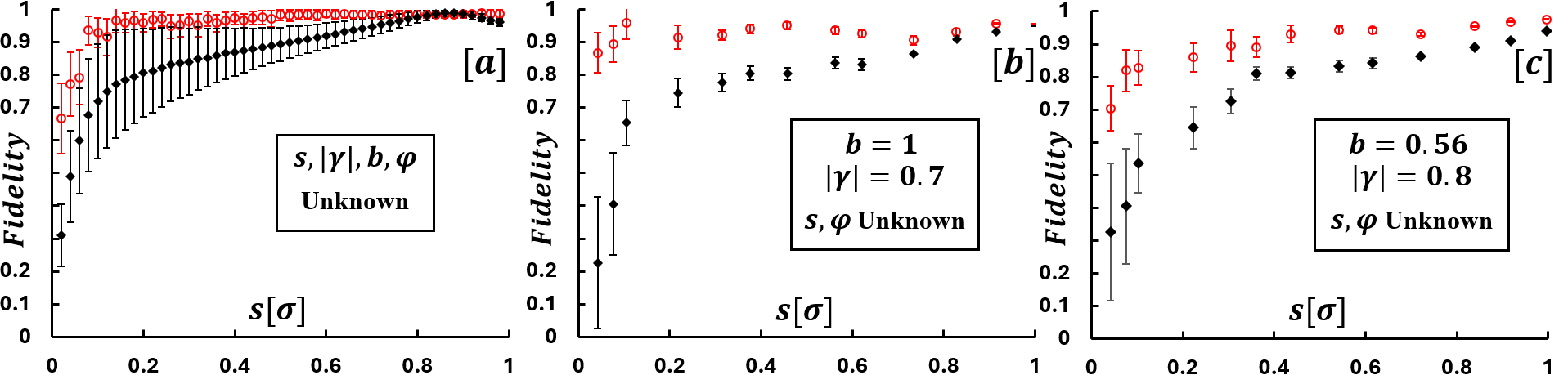}
\caption{Superresolution fidelity comparison of the parameter-decoupling physics-informed CNN model (red points) with the direct "train-and-predict" ML model (black points). (a) represents the simulated case when all four parameters ($s$, $|\gamma|$, $b$, $\varphi$) are unknown and randomized, same as Fig.~\ref{multiple-parameter noise} (d). (b) shows the case for balanced ($b=1$) and partially coherent ($|\gamma|=0.7$) experimental testing data, same as Fig.~\ref{experiment} (g).   (c) illustrates the case for unbalanced ($b=0.56$) and partially coherent ($|\gamma|=0.8$) experimental testing data, same as Fig.~\ref{experiment} (h).}
\label{Fprogram}
\end{figure*}

To assess the value of the physical representation, we compare the proposed CNN with a direct image-to-separation model trained and tested on the same images. The direct model learns the image--separation correspondence without the explicit same-parity contrast construction. Matching the data exposes both approaches to the same source-parameter distributions and image imperfections. Figures~\ref{Fprogram} (a)-(c) reproduce the fully randomized synthetic test of Fig.~\ref{multiple-parameter noise}(d), the balanced partially coherent SLM test of Fig.~\ref{experiment}(g), and the unbalanced partially coherent SLM test of Fig.~\ref{experiment}(h), respectively. Red points show the parameter-decoupling-informed estimator and black points the direct estimator. Across these tests, the physical representation provides significantly higher fidelity, particularly at small separations.

Using $F=0.8$ as a common comparison level, the supplied results indicate approximate reductions in the separation threshold by factors of 4, 15, and 9 in panels (a), (b), and (c). For example, panel (c) places the direct estimator's crossover near $s=0.37\sigma$ and the informed estimator's crossover near $s=0.04\sigma$. This comparison supports the practical benefit of the contrast-informed representation for the tested baseline.

%It does not isolate architecture, parameter count, optimization, and representation as separate causes. A capacity-matched ablation and comparisons with likelihood-based estimators and other learned reconstructions would further quantify the origin and generality of the improvement. 

%In particular, the results are not a direct performance comparison with Digital-SMLM or device-agnostic reconstruction \cite{jia2024digitalsmlm,vasinka2026stars}, which address different datasets and evaluation protocols.

\section{Conclusion and discussion}
We have developed parameter-decoupling-informed learning for quantitative estimation of two-source separation from recorded intensity images. The learning representation incorporates same-parity HG-mode log-probability contrasts, in which coherence, brightness imbalance, and relative phase cancel analytically, providing a physical constraint that guides separation inference. Synthetic and laboratory tests demonstrate sub-Rayleigh separation estimation of at least 14 times lower than the conventional resolution limit. Application to archival galaxy images further yields separation estimates consistent with previous analysis.

%It does not replace an optical mode sorter or establish that modal measurement outcomes are uniquely recoverable from arbitrary intensity images.

The synthetic tests assess source uncertainty, pixel perturbations, centroid offsets, rotations, training-set size, and sampling. Laboratory images generated with an MZ interferometer and an SLM demonstrate transfer beyond simulations. Under the partially coherent, unbalanced SLM condition, the reported separation-estimation fidelity exceeds 0.82 for $s\gtrsim0.075\sigma$. Comparisons using the same training and testing images show higher small-separation fidelity than the tested direct-learning baseline. The archival NGC~4486B application gives mean separations of 9.3--13.6~pc across the reported training configurations, consistent with the earlier approximately 12-pc estimate.

These findings concern model-based inference from information already captured by a camera. Quantum limits associated with optical spatial-mode measurements provide useful physical benchmarks, but do not establish quantum-limited performance for the image-only estimator. The learned representation neither replaces an optical mode sorter nor implies that modal measurement outcomes can be uniquely recovered from arbitrary intensity images. Nevertheless, incorporating the parameter-decoupling relation enables more effective use of the information retained in the image, substantially improving separation estimation in the deep sub-Rayleigh regime under the tested conditions.

%Broad randomized training can improve robustness within the tested domain, but does not guarantee reliable estimation across all nuisance-parameter values. 

%Its performance remains subject to finite photon budgets, PSF uncertainty, source degeneracies, and differences between training and application data.

The laboratory results and archival application play complementary roles: the former provide controlled reference separations, while the latter demonstrates retrospective use on an astronomical object with a more complex source structure. Extending the approach to calibrated photon-counting models, realistic telescope PSFs, extended sources, and uncertainty-aware outputs would strengthen its use for quantitative science. Within the conditions tested here, the central result is that a separation-specific, parameter-decoupled physical representation improves learning from overlapping source images and offers a practical route to reanalyzing existing optical data.\\

\noindent{\bf Acknowledgement:} We acknowledge financial support from NSF grants PHY-2316878 and PHY-2514953.

%, and Chin Ying Foundation

\noindent{\bf Disclosures:} The author declares no conflicts of interest.

\noindent{\bf Data Availability} The main data supporting the findings of this work are available within the article and its Supplementary Material. Additional data are available from the corresponding author upon reasonable request.

\bibliography{reference}

\end{document}